# Nuclear hyperpolarization of $^3$He by magnetized plasmas

A. Maul[1], P. Blümler[1], P.-J. Nacher[2], E. Otten[1],
G. Tastevin[2], and W. Heil[1*]

1. Institute of Physics, University of Mainz, 55128 Mainz, Germany
2. Laboratoire Kastler Brossel, ENS-PSL University, CNRS, Sorbonne Université, Collège de France, 24 rue Lhomond, 75005 Paris, France

*)   corresponding author: wheil@uni-mainz.de

**Abstract:**

We describe a method to hyperpolarize $^3$He nuclear spins at high magnetic fields (4.7 Tesla) solely by a magnetized plasma. The conditions for such a magnetized plasma are fulfilled when the mean free path of the free electrons is much larger than their gyration radius in the rf gas discharge. Investigations are carried out in the 1–15 mbar pressure range with rf excitation at ~100 MHz. Quantitative NMR measurements show that for different cell sizes and $^3$He densities nuclear polarizations in the range 1% to 9% are observed. We explain this phenomenon by an alignment-to-orientation conversion mechanism in the excited $2\,^3P$ state of $^3$He which is most efficient when the Zeeman and the spin-orbit energies are comparable. The method appears as a very attractive alternative to established laser polarization techniques (spin exchange or metastability exchange optical pumping). Application to $^3$He nuclear magnetometry with a relative precision of $10^{-12}$ is demonstrated.

# 1. Introduction

Metastability exchange optical pumping (MEOP) is an efficient method to create hyperpolarization in $^3$He, i.e. nuclear orientation which is far beyond the equilibrium polarization that can be reached at the lowest temperatures and in the highest magnetic fields available at the moment. The method, developed by Colegrove, Schearer and Walters [1], is an extension of ordinary optical pumping [2,3] where the light-induced atomic orientation is directly created in the ground state. In $^3$He MEOP operates on the excited metastable $2\,^3S_1$ state produced in a plasma discharge sustained in the gas. MEOP is usually performed in low magnetic field up to a few mTesla and efficiently operates at low pressure of order 1 mbar where nuclear polarization of up to 90% have been reported [4]. The hyperfine interaction provides the physical mechanism for the polarization transfer from the polarized light to the $^3$He nuclei. Polarization is ultimately transferred to the ground state through metastability exchange collisions. MEOP can still be used up to several Tesla and yields high polarizations even at elevated gas pressures, in spite of large hyperfine decoupling at high magnetic fields. In a recent article [5] the physics and technology of producing large quantities of highly spin-polarized $^3$He nuclei using MEOP is reviewed.

There is a rather large domain of possible applications of hyperpolarized (HP) $^3$He ranging from polarized targets for nuclear and particle physics [6,7], neutron spin filter [8-10], contrast agent in lung MRI [11,12] to measurements and monitoring of magnetic fields [13-15]. For the latter we have demonstrated that a $^3$He nuclear magnetometer is able to measure high magnetic fields ($B > 0.1$ Tesla) with a relative accuracy of better than $10^{-12}$ [13]. Our approach is based on the measurement of the free induction decay (FID) of HP-$^3$He following a resonant radio frequency pulse excitation (i.e., a pulsed nuclear magnetic resonance (NMR) experiment). The measurement sensitivity can be attributed to the long coherent spin precession time $T_2^*$ being of order of minutes which is achieved for spherical sample cells in the regime of "motional narrowing" where the disturbing influence of field inhomogeneities is strongly suppressed.

Performing such experiments routinely in our lab we observed that NMR-signals were detected even when the pumping laser was turned off accidentally. Although they were much weaker than those obtained with laser optical pumping these unexpected signals were evidence of considerable hyperpolarization (large out-of-equilibrium nuclear polarization) that ought to stem from striking the discharge (signals decayed and vanished if the discharge was turned off as well). In the context of $^3$He optical pumping experiments, Carver and coworkers have also observed discharge-induced nuclear polarization of ground state $^3$He atoms in rf discharges [16]. The reported polarization, obtained at 3.1 amagats gas density in 1 Tesla field, was about 4 times higher than the Boltzmann equilibrium polarization and had a positive or negative sign depending on the type of rf excitation (intermittent or continous discharge). The authors tentatively attributed this phenomenon to Overhauser polarization by distinct saturated species present in the plasma. Later, McCall and Carver [17] reported further investigations at lower number density (2.7 – 46.6 mbar $^3$He gas pressure) for field strengths up to 0.8 Tesla and various types of gas excitation. They reported a maximal enhancement factor of 2200, corresponding to 0.06% nuclear polarization at 0.1 Tesla, and a change in sign from positive to negative as field strength was increased (with a zero crossing at around 0.45 Tesla). They put emphasis on experimental features that would suggest that the metastable triplet atoms are involved both in the transfer of nuclear polarization to the $^3$He ground state and in the polarization enhancement process.

Our studies are performed at higher field strength (a few Tesla), and such a mechanism cannot account for our results, because the polarization levels we measure significantly exceed even that of thermally polarized electrons or paramagnetic atoms. Since production of HP $^3$He gas without lasers will be of great practical advantage, particularly for use in magnetometry [13] and because the origin of the observed effect seems not yet clearly established, we have investigated the influence of the operating conditions on achieved polarizations and buildup rates. Here, we report on NMR measurements performed on spherical $^3$He gas samples at 4.7 Tesla for different cell volumes, filling pressures, or rf excitation levels. The experimental setup is described in Section 2. The collected data are presented in Section 3. A first demonstration of application to high-field magnetometry is discussed in Section 4. The details of the determination of the absolute polarization are described in Appendix A, and the explanation of the effect via an alignment-to-orientation conversion mechanism (AOC) can be found in Appendix B. We suggest to use the acronym PAMP, for **P**olarization of **A**toms in a **M**agnetized **P**lasma, to describe the method that allows to obtain nuclear hyperpolarization solely from a gas discharge in which the mean free path of the electrons is large compared to their cyclotron radius.

## 2. Experimental

Figure 1 shows a schematic drawing and photograph of the experimental setup. The sample consists of a spherical glass cell which is filled with a few mbar of pure $^3$He. Several cells with inner diameter $\varnothing_{ID}$ were used (8 mm $< \varnothing_{ID} <$ 20 mm), all blown from standard Pyrex glass with a wall thickness of ca. 1 mm. Each cell was successively cleaned with Mucasol[1] and rinsed with distilled water, evacuated, baked out, and finally filled with the desired $^3$He pressure before it was sealed off by a torch. Cells were mounted inside a NMR-probe coil (see Fig. 1b) and placed inside a superconducting magnet at 4.7 Tesla (homogeneity ca. 1 ppm/cm). The remaining sealing-stem of the cells was always oriented perpendicular to the direction of the magnetic field in order to reduce field gradients across the sample volume originating from magnetic susceptibility mismatch [13]. For MEOP experiments the discharge coil was a solenoid and the laser was shone on the sample through its inner core. Otherwise the discharge coil was tightly

---

[1] Mucasol is a trademark of Merz GmbH & Co. KG: universal cleaning agent for labware and instruments made of glass

wound onto the glass cell. In both cases, the discharge coil axis was oriented parallel to the magnetic field.

The discharge coil was part of a serial LC circuit carefully tuned and matched prior to each experiment. Coil dimensions and available rf-capacitators typically yield resonance frequencies in the 100-120 MHz range produced by a sine-generator, amplified, and fed into the LC circuit. Matching conditions (minimal reflected rf signal, purely resistive impedance $R = 50 \, \Omega$) were constantly monitored using a bidirectional high-power rf coupler (cf. Fig. 1a).

NMR excitation and detection was performed with either a Helmholtz (for MEOP experiments) or a solenoidal rf coil, tuned to the Larmor-frequency of $^3$He, $f_L$, ($f_L$ =152.26 MHz) and oriented perpendicular to the magnetic field. The entire experiment was controlled by a KEA spectrometer[2] that managed the NMR excitation and signal acquisition as well as the gating of the discharge via TTL control of the rf amplifier. Typical NMR-acquisition parameters were: flip-angle = 90°, pulse length 40 – 50 µs, dwell times 0.1 – 1 ms. If not mentioned otherwise, the FID signal from a single NMR-excitation was recorded. The initial amplitude, $S$, of the FID signal was used as a measure of the magnitude of the nuclear polarization. Absolute polarization values were inferred from the NMR sensitivity factor derived from calibration measurements (Appendix A).

The optical part of the experiment was a standard MEOP-setup, as described in [13]. It includes a 1083 nm laser source for excitation of the $2^3$S-$2^3$P transition as well as a circular polarizer and optical elements for light beam control. An infrared photodiode (sensitivity range: 850 nm – 1070 nm) located next to the $^3$He cell monitors some amount of fluorescent light emitted by the discharge. In the present work, the photodiode signals were principally used as indicators of the discharge brightness. Here we pragmatically report the rf excitation level in terms of the effective power, $p_{eff}$, which is dissipated in the discharge circuit and helium cell. To this aim, we use the measurements of forward, $U_f$, and reflected, $U_r$, rf voltages to infer:

$$p_{\text{eff}} = (U_f - U_r)^2 / 2R \tag{1}$$

The build-up[3] of the nuclear polarization signal $S(t)$ in the $^3$He plasma detected via the monitored FID signal may generally be described by a single exponential growth rate, $\Gamma$, and an asymptotic value, $S_\infty$, such that :

$$S(t) = S_\infty \cdot (1 - \exp(-\Gamma \cdot t)) \tag{2}$$

---

[2] Magritek, Unit 3, 6 Hurring Place Newlands Wellington 6037 NEW ZEALAND
[3] The signal amplitude of the nuclear polarization was probed at different times in consecutively repeated experiments each starting with initial polarization $P(t = 0) = 0$.

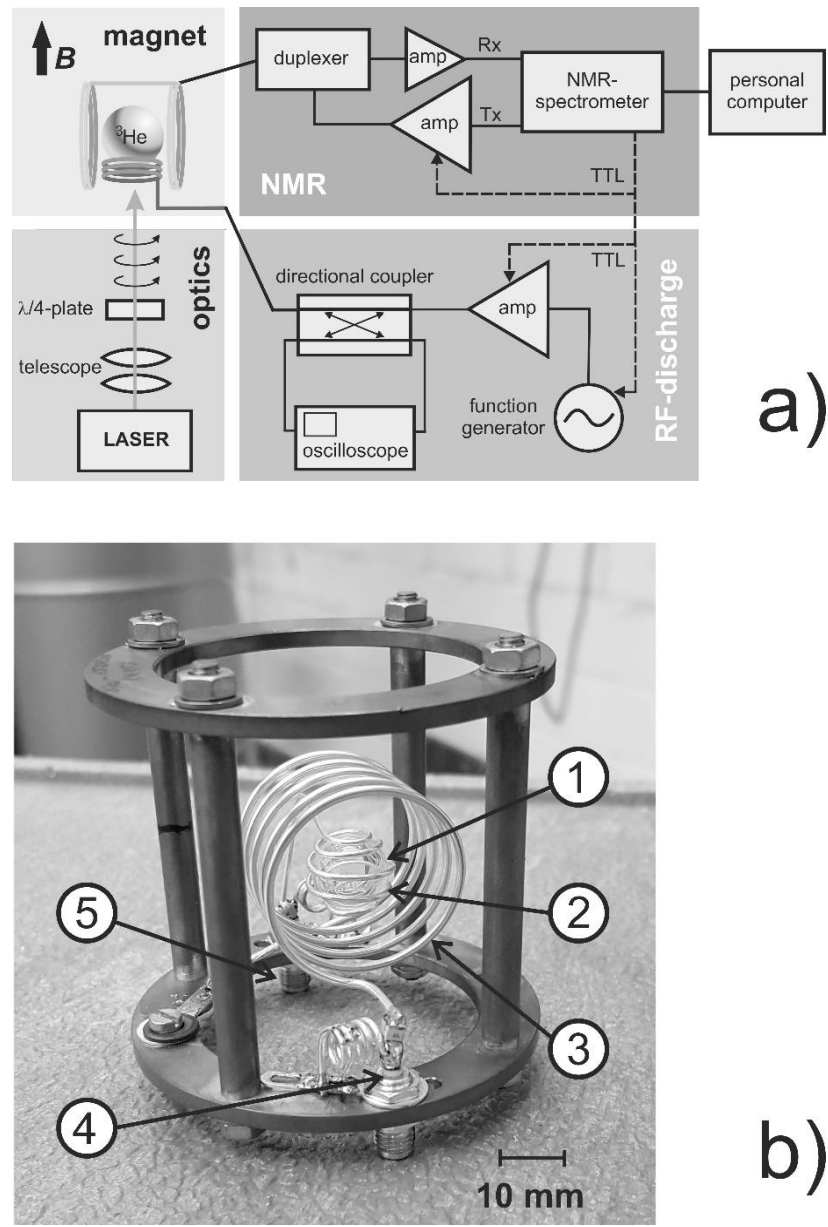

Fig. 1: a) Schematic drawing of the experimental setup. The sample is immersed in a strong and homogeneous magnetic field, $B = 4.7$ Tesla. The depicted probe configuration, used for MEOP, includes a solenoidal discharge coil (holding the cell) and a Helmholtz NMR coil pair. It is connected to three experimental parts: (*optics*) circularly polarized light of 1083 nm provided by a 2W Yb-doped fiber laser[4]); (*RF discharge*) a discharge circuit for plasma ignition; (*NMR*) NMR-excitation and detection. The NMR spectrometer controls and synchronizes the entire setup via a data connection to a PC. b) Photograph showing the probe configuration for PAMP: 1) Spherical glass sample with $^3$He, 2) discharge coil wound on the gas container, 3) NMR solenoidal coil, 4) NMR-coil connector, 5) discharge coil connector.

---

[4] Keopsys, 2 Rue Paul Sabatier, 22300 Lannion, France

## 3. Discharge polarization results

This section gives a first quantitative survey of the various dependencies of the gas discharge polarization on dedicated parameter settings together with the determination of the orientation of the $^3$He nuclear spins.

### 3.1 Sign of the $^3$He nuclear polarization

The nuclear polarization, $P$, is defined as

$$P = \frac{N_+ - N_-}{N_+ + N_-}, \qquad (3)$$

where $N_+$ and $N_-$ are the population numbers for the two nuclear spin states, $m_I = \pm\frac{1}{2}$ of $^3$He ($I = \frac{1}{2}$). Unambiguous determination of the orientation of the nuclear spins produced by the rf excitation can be obtained by comparison with that achieved by MEOP. This comparison is particularly easy to perform at 4.7 Tesla thanks to large Zeeman energy splittings between magnetic eigenstates and strong differences in resonance line positions in the 1083 nm absorption spectrum of $^3$He; see Fig.1 in [18].

We have selected two strong absorption lines, the so-called $f_4^+$ - and $f_2^+$ -transition lines of $^3$He, which lie within the 120 GHz broad tuning range of our laser [19] and are well resolved at room temperature (the atomic Doppler FWHM, 2 GHz, is small compared to the $f_4^+ - f_2^+$ line splitting of 9.1 GHz). They both belong to 1083 nm absorption spectrum of $^3$He for the same circular light polarization but yield nuclear spin polarization with opposite signs [20]: the $f_2^+$ -transition depletes $N_-$ (hence, creates $P > 0$) and the opposite is true for pumping via the $f_4^+$ -line.

Two experiments have been performed in which $^3$He gas was maintained under constant rf excitation (except during NMR measurements) and the laser, tuned to one resonance line ($f_2^+$ or $f_4^+$) was shone onto the sample for a 20 s period of MEOP, then blocked. The results are shown in Fig. 2. In both cases, the NMR signal rapidly increases during the MEOP period, towards a finite asymptotic value, and starts to decay when the laser is blocked. For $f_4^+$ -pumping (Fig. 2a) the signal amplitude monotonically decays towards a finite and smaller asymptotic value. For $f_2^+$ -pumping (Fig. 2b) the signal amplitude decays, reaches a null value, then grows towards an (also smaller) asymptotic value. The change in sign is indicated by the kink at the point of zero polarization. The fact that the asymptotic NMR signals diverge by about a factor of two is to be attributed to the difference in rf discharge powers used during both examinations. From theses observations, we conclude that PAMP induces *negative* $^3$He nuclear polarization in 4.7 Tesla, as does $f_4^+$ - pumping.

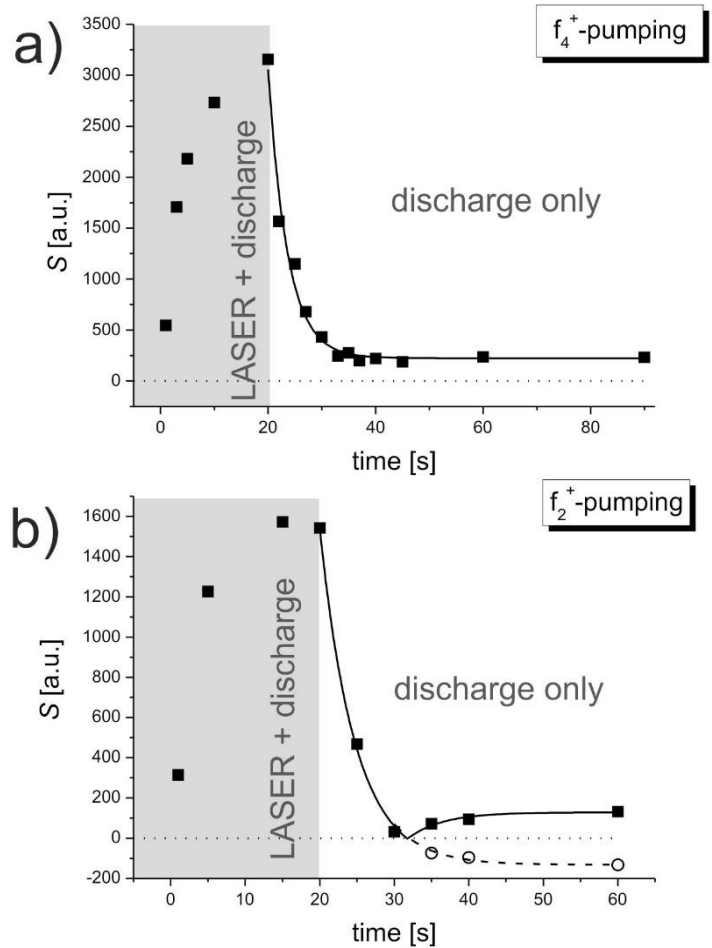

Fig. 2: Development of the NMR-signal of $^3$He in an experiment where polarization was built up by MEOP in the first 20 s (gray shaded area); thereafter the laser was blocked while the rf discharge was kept burning. The magnitude of the nuclear polarization (black solid symbols) was probed at different times in consecutively repeated experiments. For this the discharge was switched off and a NMR experiment was performed. a) Polarizing with the laser tuned to the $f_4^+$-transition and b) to the $f_2^+$-line. Both experiments were done at 4.7 Tesla on a spherical sample (inner diameter: 8 mm, pressure ca. 1 mbar). The full symbols represent the measured initial amplitudes of the NMR signal while the open symbols in b) correspond to the opposite (mirrored) values. The curves describing the $^3$He discharge-relaxation after blocking the laser are mono-exponential decays.

## 3.2 Dependence on discharge power

By increasing the applied rf power ($p_{\text{eff}}$, as obtained with eq. (1)) a strong growth of the NMR-signal was observed. It was also noticed that not all power must have been transferred to the plasma. First of all the electrical losses cause heating of the tank circuit at higher currents. This will then detune the tank circuit driving it out of the resonance conditions. Besides that the power dissipation in capacitively (or inductively) coupled rf discharges has to be considered; this has been studied, e.g., in [21]. The properties of capacitively/inductively coupled discharges are strongly influenced by the discharge intrinsic structure. It consists of a positive column-like discharge volume, the glow space or "bulk" and two specific interaction regions between the bulk and the dielectric walls in front of the field-supplying electrodes, the "sheaths."[5] The power dissipation in the sheath regions [22][6] is one of the main loss processes and generally increases with increasing rf-current. Thus, the characterization of rf discharge conditions by the transmitted rf power measurements is weakly relevant since only a small fraction of the total measured rf power is related to the electron heating process which governs the electron energy probability function (EEPF) in the luminous bulk plasma [23]. Therefore the luminous intensity of the bulk plasma was monitored by means of a photo-diode. Figure 3a shows the dependence of the plasma light intensity versus the effective discharge power which can be described by an exponential approach towards a saturation value and has been used as a reference for the actual plasma intensity.

To investigate the influence of discharge power on the polarization build-up, NMR-measurements were made about 1 s after the discharge was stopped. The rf power, $p_{\text{eff}}$, was varied in consecutive runs from ca. 1 to 50 W. Figure 3b shows its influence on the respective build-up curves of the observed $^3$He NMR-signals, i.e., a general increase of both polarization build-up rate ($\Gamma$) and saturation polarization ($S_\infty$) with increasing rf-power. The situation becomes clearer (cf. Fig. 3c) when these curves are fitted with eq. (2) and the fit-parameters $S_\infty$ and $\Gamma$ are plotted versus the plasma intensity as measured by the photo-diode in Fig. 3a. This removes the strong non-linearity between the applied electric power and the intensity of the induced plasma, the bulk plasma. Of course, there are insufficiencies in using the $U_D$ signal to describe the bulk plasma: self-absorption, emitted spectrum changes as the discharge power is increased, finite spectral range of photodiode, etc. The obvious stronger increase of the $S_\infty$ and $\Gamma$ values (outliners) at the highest discharge power (cf. Fig.3c) may be attributed to a not one-to-one assignment of the actual discharge power in the bulk plasma as measured by the photo-diode.

---

[5] A sheath layer is several Debye lengths thick. The value of this length, $\lambda_D = \sqrt{\varepsilon_0 \, k_B \, T_e / (n_e e^2)}$, depends on various characteristics of the plasma (e.g., electron temperature $T_e$ and density $n_e$). In a weakly ionized gas discharge: $\lambda_D \sim 0.15$ mm.

[6] Typical electron temperatures $T_e$ in a glow discharge plasma are in the range of 1-10 eV, ion temperature and neutral gas temperature are relatively low, around 0.03 eV. However, the energy of ions bombarding the substrate can exceed $T_e$. This is due to the net positive space charge in the plasma sheath leading to potential profile that falls sharply to the local substrate potential near the boundary.

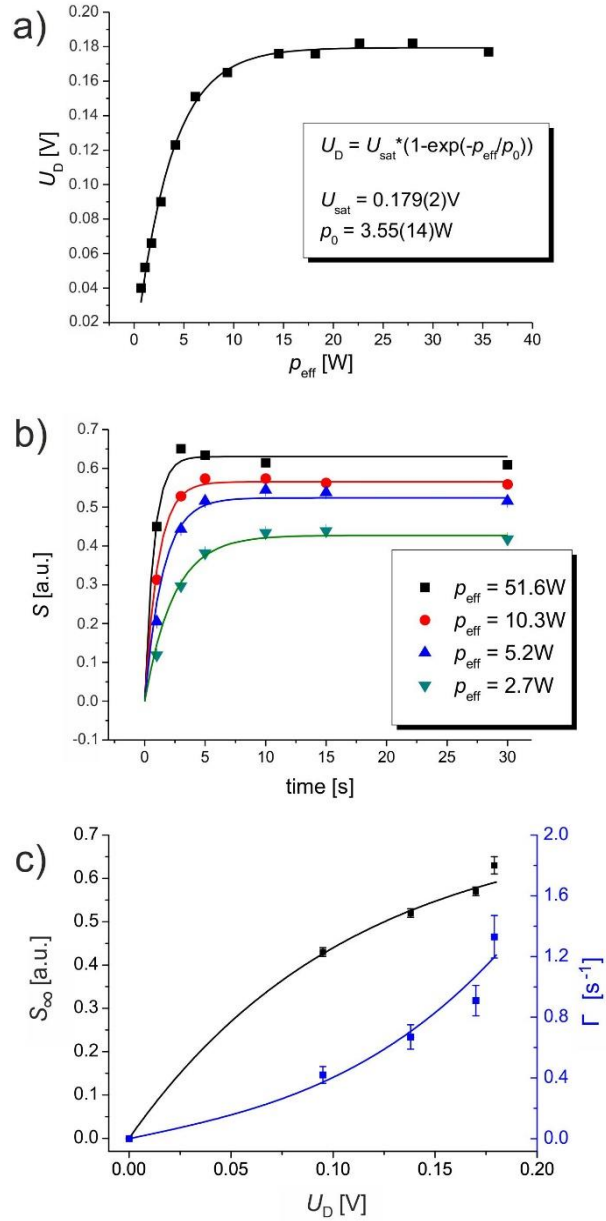

Fig. 3: a) Photo-diode voltage signal, $U_D$, versus the electrical power $p_{eff}$ applied to the discharge coil. The black curve represents an exponential fit. b) $^3$He NMR-signals versus discharge duration for four different discharge powers(2.7 W, 5.2 W, 10.3 W, and 51.6 W, from bottom to top). The solid lines are fits of eq. (2) to the data. This graph is meant to show results from a spherical cell ( $\varnothing_{ID}$ = 10.8 mm and $p_{He}$ of about 1 mbar) yielding relatively high polarization build-up rates ($\Gamma \approx 1$ s$^{-1}$). Much smaller $\Gamma$'s were observed for other sample diameters and pressures (see section 3.3) . c) Fit parameters obtained for the data in b). $S_\infty$ (black squares) and $\Gamma$ (blue dots) are plotted against the photodiode signal $U_D$, using the fit in a). When discharge is off ($U_D = 0$), no polarization signal is observed ($S_\infty = 0$) and $\Gamma$ essentially reduces to the negligible contribution from the wall relaxation rate (hence, we set $\Gamma(U_D = 0) \approx 0$). The blue solid line is a third order polynominal fit to the measured $\Gamma$-data,. The lines are essentially guide for the eyes since no explicit model is involved.

## 3.3 Influence of sample volume and ³He gas pressure

Polarization influencing parameters which can easily be varied are sample size and filling pressure. Therefore, three different types of spherical sample cells of average size $\varnothing_{ID}$ = 8.2, 10.8, and 19.2 mm were filled with four different ³He-pressures ($p_{He}$ = 1.0(5), 5.0(5), 10.0(5), and 15.0(5) mbar), i.e., a total of 12 cells were prepared. Measurements have been performed at discharge powers $p_{eff}$ in the 30–50 W range. While the discharge coils were always snugly fitted to the sample size, the NMR-detection circuit was not altered to ensure comparable NMR sensitivity. Like in Fig. 3b the amplitudes of the observed ³He NMR-signals were recorded as a function of the discharge duration. Using the mono-exponential saturation law from eq. (2) the fit parameters $S_\infty$ and $\Gamma$ were extracted. In order to reference the measured NMR signal $S_\infty$ to the corresponding polarization value $P_\infty$, a NMR signal calibration was performed with thermally polarized ³He samples. The procedure is described in Appendix A. By use of eq. (A4) the corresponding magnitude of polarization values $|P_\infty|$ could then be deduced. They are compiled and plotted as a function of the cell filling pressure in Figure 4. In all cases the discharge induced nuclear polarization of ³He lies well beyond the thermal limit of electronic Boltzmann-polarization which is $P_{th}^{e^-}$ = –1.08% at 4.7 Tesla and room temperature (293 K).

Looking at the characteristic polarization build-up rates, $\Gamma$, (see Fig. 5) there is no uniform picture in the various dependencies on pressure and sample volume except for a general decrease with pressure. Given a constant rf-power in the bulk discharge plasma one might expect $\Gamma$ to decrease with increasing sample volume and pressure according to $\Gamma \propto 1/(p_{He} \cdot V)$. A general decrease towards higher filling pressures can be observed particularly at the sample cells with the largest size ($\varnothing_{ID}$ = 19.2 mm), but the expected volume dependency is hardly pronounced. In one particular measurement, namely with the cell $\varnothing_{ID}$ = 10.8 mm we have observed a giant rise of $\Gamma$ by almost a factor 100 in lowering the pressure from 5 mbar to 1 mbar (see Fig. 5). This is associated with a drop in nuclear polarization, too. From the plasma light intensity which was also recorded we can deduce a strong dependence on the fine-tuning of the rf matching circuit which maximizes the effective power delivered to the plasma. That may explain to some extent an observed resonance-like increase of the polarization build up rate for that particular case. But also abrupt changes in the EEPF shape with a corresponding drop in the effective electron temperature and a rapid increase of the plasma density are well known in the rf discharge literature [23]. This goes along with the transition from the α- to the γ mode at given gas pressures and rf discharge strength[7]. In the α-mode (Joule heating), electron and ion motion in the plasma body are collisionally dominated with the plasma's spatial distribution controlled by ambipolar diffusion and collisional electron heating is the main rf power dissipation process. In the γ mode, secondary electrons[8] (born at the rf electrodes due to ion bombardment) and other electrons (due to electron avalanche or ion-ionization in the sheaths) accelerate in the rf sheaths towards the plasma where they perform intensive ionization and excitation.

At strong magnetic fields, the magnetic confinement (see Appendix B) drastically changes the ionization balance and the plasma's radial distribution (particularly inhomogeneous at higher gas pressures, for $p_{He}$ > 30 mbar [24]). In that case, the light-emitting plasma is generally non-uniform and mostly located close to the walls of the cell. Moreover, the nature of the rf discharge is also deeply modified by cell geometry (cell size, thickness of dielectric walls, etc.) and rf excitation frequency [22,25].

In summary, these rather complex interrelationships do not allow us to explain on a quantitative basis the experimentally observed dependencies of $|P_\infty|$ and $\Gamma$ on sample volume and $^3$He gas density (cf. Figs. 4 and 5). They certainly would require a deeper investigation on capacitively or inductively driven rf discharges, in particular the monitoring of the evolution of the spatial density distribution of the 2 $^3$S metastable states and the role of the EEPF integral-based quantities as the plasma density $n_e$ and the effective electron temperature $T_e$ (see Appendix B). However, in view of the application of PAMP to high-precision magnetometry (which requires small spherical cells) such an investigation seems both difficult and unnecessary.

---

[7] In [23] transitions can occur in the pressure region $\Delta p$ < 5 mbar depending on the current density of the rf gas discharge.
[8] In an electrodeless discharge, this should not be relevant.

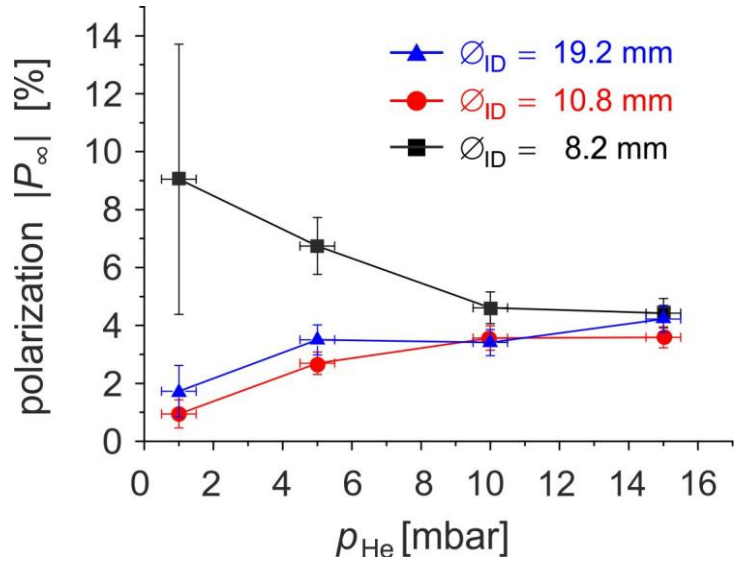

Fig.4: Measurements of absolute polarization $|P_\infty|$ produced by PAMP in various spherical samples of different $^3$He-pressures and three different cell diameters (black squares: $\varnothing_{ID} = 8.2$ mm), red dots: $\varnothing_{ID} = 10.8$ mm, blue triangles $\varnothing_{ID} = 19.2$ mm). The absolute polarization was determined by the procedure described in Appendix A. The increase of the error bars towards lower gas pressures essentially results from the overall pressure uncertainty of ±0.5 mbar (cf. eq. (A4)). The measurements were conducted in a magnetic field of $B = 4.7$ Tesla.

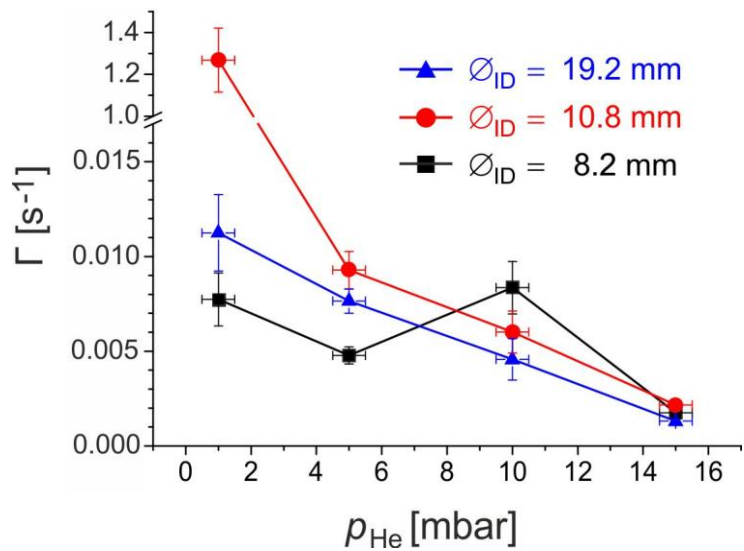

Fig.5: Measurements of polarization build-up rate $\Gamma$ for various gas pressures and sample diameters $\varnothing_{ID}$ (black squares: 8.2 mm, red dots: 10.8 mm, blue triangles: 19.2 mm). The notably larger (by a factor 100) build-up rate $\Gamma$ measured in the $\varnothing_{ID} = 10.8$ mm cell is presumably due to the onset of a different discharge mode.

## 4. PAMP-based Magnetometry

Since a potentially important use of the PAMP-effect is to design simple and extremely sensitive magnetometers, we now demonstrate this application with two experiments performed with 1 mbar gas samples. The first sample was contained in a $\varnothing_{ID}$ = 8.2 mm glass sphere with a wall thickness of ca. 1 mm, and a short stem, selected among the set of $^3$He cells used for the NMR measurements reported in Section 3. The $^3$He sample has been polarized by rf excitation with $p_{eff}$ = 39 W up to saturation polarization[9]. Figure 6a shows the recorded FID signal following a single 90° NMR pulse. The high *SNR* of 1050 (referenced to a bandwidth, $f_{BW}$, of 1 Hz) clearly demonstrates the efficiency of polarizing the $^3$He via PAMP. In this experiment, a pronounced stem significantly shortened the signal lifetime ($T_2^*$ = 1.3 s) as a result of a non-spherical susceptibilty distribution that causes static field inhomogeneities [13]. The second sample was contained in a quartz cell of almost perfect sphericity and similar diameter ($\varnothing_{ID}$ = 8.0 mm) designed and used for $^3$He magnetometry [26]. In contast to the first sample cell, the FID signal, shown in Fig. 6b, had a much longer lifetime ($T_2^*$ = 40 s). The lower *SNR* of the FID recording (ca. 100 @ $f_{BW}$ =1 Hz) measured with this cell may be attributed to the relatively thick dielectric container wall of 2 mm hampering the rf excitation of the gas. Besides, it was recognized that the gas discharge showed a slightly different color which is an indication of gas contamination stemming from impurities desorbed from the walls during the discharge process. Gas impurities quench the density of metastable $^3$He atoms and thus may reduce the PAMP efficiency (see Appendix B).

Following the data treatment presented in [13], the magnetometers' sensitivity across the respective $T_2^*$ time interval is $\delta B/B = 2.5 \cdot 10^{-12}$ for Fig. 6a and $\delta B/B = 1.1 \cdot 10^{-12}$ for Fig 6b. Here two very different data result in a very similar sensitivity because the sensitivity scales $\propto SNR \cdot T_2^*$. The obvious advantage of a much longer $T_2^*$, however, pays off when monitoring and controlling a field. Therefore, further improvement of the PAMP-sensitivity requires thin walled cells of perfect sphericity and presumably very high gas purity even at the elevated cell temperatures.

---

[9] The characteristic polarization build-up time was $\tau_{pol} \approx$ 130 s. In case of the spherical quartz cell (second sample) we measured $\tau_{pol} \approx$ 3 s .

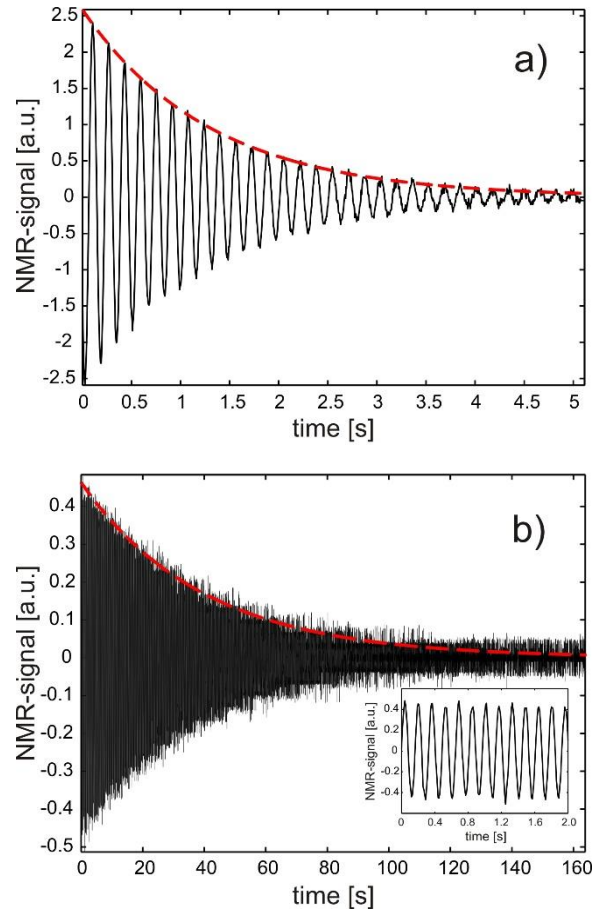

Fig.6: FID signal (real-part only) measured at low beat frequency $f_b = f_L$ -152.26 MHz from two samples polarized via PAMP at 4.7 Tesla: a) cell with stem and 1 mm thick walls ($\varnothing_{ID}$ = 8.2 mm, $p_{He}$ = 1 mbar); 1k data were acquired with a dwell time of 5 ms, b) sample of high sphericity but 2 mm thick walls ($\varnothing_{ID}$ = 8.0 mm, $p_{He}$ ca. 1 mbar [26]); 16k data were sampled with a dwell time of 10 ms. The insert shows the first 2 seconds of the signal for direct comparison with a).The red dashed line indicates a mono-exponential fit to the data in order to determine $T_2^*$ which is 1.28 s for a) and 39.7 s for b). Note the different scales of the time axes.

## 5. Discussion and Outlook

The main objective of the present article is to report, first, on the observation of $^3$He hyperpolarization solely due to plasma excitation (PAMP) in presence of a high magnetic field (4.7 Tesla) and, second, on the measurement of ground state nuclear spin polarization a few times higher than the thermal Boltzmann limit of free electrons. The PAMP-mechanism is described in terms of an alignment-to-orientation conversion in helium (cf. Appendix B).

Complementary tests have been made to provide a quick overview of the dynamic range of PAMP. Using available high field magnets and NMR systems, we have checked that significant NMR signals could be obtained at various field strengths and temperatures. Under each of such conditions no attempt was made to accurately quantify or optimize the observed signal. We operated in 1.5, 7, and 11.7 Tesla magnets using a blown spherical cell with $\varnothing_{ID} \approx 20$ mm and 1.2 mbar filling pressure. Relying on local informations for the NMR calibration, the evaluated nuclear polarization values fall within the $0.1\% < |P_\infty| < 1\%$ range.

PAMP tests were performed at 77 K and around 10 K in a 4 Tesla magnet (used for Penning trap experiments in Heidelberg) using a 1.5 mbar $^3$He cell of similar size. The actual temperature of the gas was not measured. Moderate rf power (< 3W) was used to limit thermal load. NMR amplitudes could not be calibrated nor quantitatively compared. But as measured *SNR* of the FID signal we reached 670:1 at 77K and 90:1 at 10K in a bandwidth of 1 Hz (for comparison see Fig. 6).

Further tests of PAMP have been performed in a $^4$He-rich isotopic mixture and in pure $^{129}$Xe gas. They were both motivated by prior MEOP experiments. The addition of $^4$He in MEOP cells is known to yield higher nuclear $^3$He polarizations and shorter build up rates, provided that optical pumping selectively operates on $^4$He metastable state atoms [27]. The higher probability of light absorption by the even isotope contributes to make this indirect pumping method more efficient in many situations (in spite of the needed transfer of orientation to the metastable state $^3$He atoms, by metastability exchange collisions between the optically-pumped metastable state $^4$He atoms and ground state $^3$He atoms, as a first step). Similarly, addition of $^4$He might be advantageous for PAMP if metastable state atoms play a key role in the polarization process. This is apparently not the case as tested with a $\varnothing_{ID} \approx 20$ mm gas cell filled with 3.0 mbar of $^3$He and 7.2 mbar of $^4$He at 4.7 Tesla. However, we cannot exclude that a higher density of impurities in the test sample of isotopic gas mixture may have limited PAMP efficiency.

No PAMP signal has been detected in the $^{129}$Xe test cell ($\varnothing_{ID} \approx 20$ mm), filled with low pressure $^{129}$Xe gas around 2 mbar, at room temperature and 4.7 Tesla. We believe that NMR sensitivity would have allowed detection of nuclear polarization if it were comparable to (or moderately smaller) than that achieved in $^3$He. The absence of sizable signal is in line with the negative result of the attempts to use MEOP for hyperpolarization of xenon: optical polarization of $^{129}$Xe metastable state atoms is successful, but ground state nuclear polarization systematically fails to be detected [28,29].

In the context of precision magnetometry, small cells are used as field probes. Detailed quantitative experimental proof of the underlying physics and analysis of the effects of PAMP within such field probes seem very difficult and are not planned. Some technical improvements will still be performed. Already in its present status precision magnetometry obviously benefits from the PAMP-effect, because the experiment can be compacted and miniaturized (thanks to dispensable optical components). A comprehensive study of the field dependence of PAMP-efficiency would be very interesting. From eq. (B4) we may expect that the efficiency will drop significantly when the electron mean free path λ approaches the gyration radius $r_c$. However, from the work of Carver et al. [16,17] we expect this drop to occur below 0.1 Tesla. The limit to PAMP efficiency at high magnetic fields will probably be set by full (fine and hyperfine) magnetic decoupling in the excited states [30].

In conclusion, our experimental findings seem very encouraging but obviously call for further work. Systematic investigations are needed to establish the full potential of PAMP. Theoretical work is also highly desirable and would facilitate optimization and exploitation of the method.

## 6.  Acknowledgements

We are thankful to Rainer Jera for providing innumerous glass cells on demand. Financial support by the Deutsche Forschungsgemeinschaft (DFG) under He2308/16-1 and by the cluster of excellence PRISMA "Precision Physics, Fundamental Interactions and Structure of Matter", is also greatly acknowledged.

# Appendix A: NMR signal calibration with thermally polarized $^3$He samples

In order to determine the absolute polarization of $^3$He from NMR-data it is possible to reference the measured amplitudes versus the known thermal nuclear polarization:

$$P_{th} = \tanh\frac{\hbar\gamma B}{2k_B T}, \tag{A1}$$

where $\hbar$ is the Planck constant, $\gamma$ the gyromagnetic ratio of $^3$He ($\gamma_{He}/2\pi = -32.434$ MHz/Tesla [31]) and $k_B$ the Boltzmann constant. It is small ($10^{-5}$ at room temperature and 4.7 Tesla, typically; see Table A1), hence, high gas density and fast signal averaging (of $\bar{n} \approx 1000$ FID signals) are needed for accurate reference measurements. Sealed glass spheres of different volumes, filled with relatively high $^3$He pressures (typically 0.5 bar) were prepared. Since the bulk longitudinal nuclear relaxation time $T_1$ can amount to hours in pure $^3$He, it needs to be shortened for fast signal averaging. Relaxation can be much faster if molecular oxygen is added, but about 3 bars of $O_2$ are required to obtain $T_1$ as short as 1 s [32]. Therefore, the thermal decomposition of an inorganic peroxide was used to release $O_2$ after sealing. We used strontium peroxide which decomposes at 215°C according to:

$$2\,SrO_2 \xrightarrow{\Delta T} 2\,SrO + O_2. \tag{A2}$$

A stoichiometric amount of $SrO_2$ powder has been introduced in the glass spheres, prior to evacuation, filling with $^3$He, and sealing. Then, the cells have been placed into an oven at 300°C to release the oxygen.

Table A1 lists the filling pressures ($p_{He}$) and the sample volumes ($V$) of the four glass spheres used to measure the calibration coefficient, $\eta$, that relates the average (initial) FID signal amplitude, $\langle S_{th} \rangle$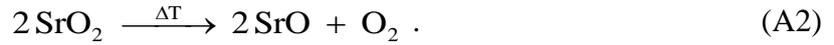, to these experimental parameters and the equilibrium nuclear polarization:

$$\langle S_{th} \rangle = \eta |P_{th}| V\, p_{He}. \tag{A3}$$

Table A1: Parameter values for the four $O_2$-doped $^3$He gas samples used for NMR signal calibration.

| Sample | #1 | #2 | #3 | #4 |
|---|---|---|---|---|
| $V$ [mm$^3$] | 711(58) | 1830(220) | 2631(280) | 3547(343) |
| $\varnothing_{ID}$ [mm]. | 11.08(30) | 15.18(60) | 17.12(60) | 18.92(60) |
| $p_{He}$ [mbar] | 495.0(5) | 495.0(5) | 650.0(5) | 495.0(5) |
| $p_{O2}$ [mbar] | $\approx 3000$ | $\approx 3000$ | $\approx 3000$ | $\approx 3000$ |
| $10^5 \times P_{th}$ ($T$ = 293 K, $B$ = 4.7 Tesla) | −1.25(2) | −1.25(2) | −1.25(2) | −1.25(2) |
| $\langle S_{th} \rangle$ [a. u.] ($\bar{n} \approx 1000$, typically) | 0.479(1) | 1.245(4) | 2.054(10) | 2.170(13) |

Figure A1 shows the compilation of the values of η inferred from eq. (A3), obtained from NMR measurements performed at fixed amplifier gain on the four $O_2$-doped $^3$He samples listed in Table A1. Within experimental error bars, the data are consistent with a constant value of the coefficient η, which shows that potential differences in filling factors (which combine coil sensitivity and rf field inhomogeneity averaged over the sample volume), are negligible.

The absolute value of discharge-induced nuclear polarization, $|P_\infty|$, asymptoticaly reached in a sealed $^3$He cell of volume $V$ and filling pressure $p_{He}$, can thus be derived from the measured signal amplitude $S_\infty$ using:

$$|P_\infty| = \frac{S_\infty}{\eta V p_{He}}. \tag{A4}$$

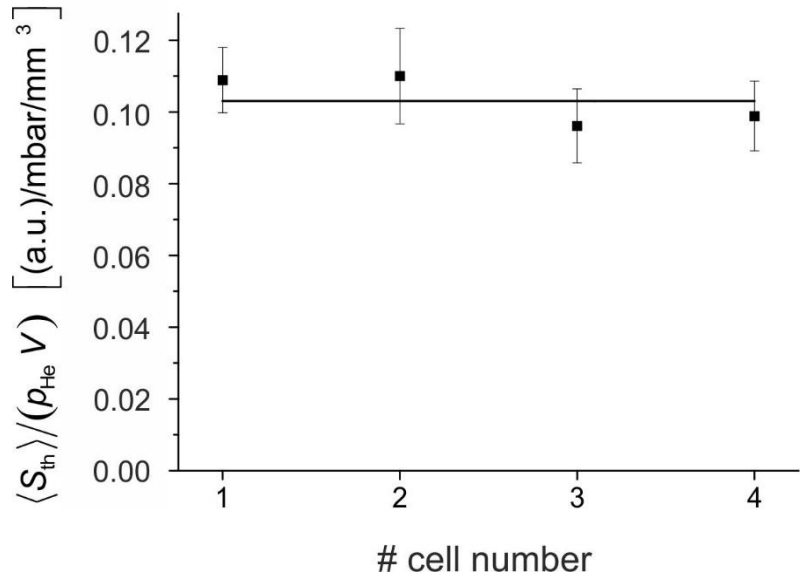

Fig.A1: Compilation of data obtained from NMR measurements performed on thermally polarised $^3$He gas, using four $O_2$-doped samples listed in Table A1. They yield the weigthed average value of the calibration coefficient introduced in eq. (A3): η = (0.103±0.005) [a.u.]/mbar/mm$^3$.

# Appendix B: Considerations on alignment-to-orientation conversion in helium

In the fields of optical spectroscopy and atomic physics, gas discharges are known as excellent media for the observation and use of unbalanced distribution of populations between energy levels, but also between magnetic sublevels. Lombardi and Pébay-Peyroula [33] showed that high frequency capacitive discharges could be used to induce alignment in a gas, as a result of an anisotropic bombardment. Fano, Lehmann, Lombardi, and others, have exhibited conditions in which the alignment could develop into orientation and proposed theoretical descriptions of the so called alignment-to-orientation conversion (AOC) processes (see [34,35] and references therein). Lombardi also proposed application of AOC in high frequency discharges for polarization of nuclear spin of $^3$He ground state atoms [36,37], but we are not aware of such experiments. In the following a first but still incomplete description of PAMP is given, in terms of selective atomic excitation by free electrons and AOC-induced nuclear polarization of $^3$He through $2\,^3P \to 2\,^3S$ radiative decay and metastability exchange collisions.

## B.1 Electronic excitation in the rf discharge

The time averaged rf power dissipated into the plasma is given by the volume integral

$$P_d = \oiiint \sigma |E_{rf}|^2 \, dV, \tag{B1}$$

where $E_{rf}$ is the amplitude of the rf electric field and $\sigma$ is the plasma conductivity given by

$$\sigma = \frac{e^2 \bar{n}_e \nu_e}{m_e (\omega^2 + \nu_e^2)}. \tag{B2}$$

Here, $\bar{n}_e$ denotes the average electron density, $\omega$ is the rf field frequency and $\nu_e$ is the electron collision frequency. The strong magnetic field affects the transverse plasma transport in the cell due to the reduction of the transverse conductivity $\sigma_\perp$ according to

$$\sigma_\perp = \frac{\sigma}{1 + \omega_c^2 / \nu_e^2}, \tag{B3}$$

where $\omega_c = eB/m_e$ is the cyclotron frequency of the electron. While the transverse component of the conductivity decreases with increasing $B$ and becomes very small for $\omega_c \gg \nu_e$, the magnetic field does not affect the longitudinal component of the conductivity $\sigma_\parallel = \sigma$. The condition $\sigma_\perp \ll \sigma_\parallel$ characterizes magnetic confinement and applies to the so called magnetized plasmas. It is met whenever the cyclotron- (or gyration) radius $r_c$ is much smaller than the mean free path $\lambda = (\sigma_{coll} \cdot n_{He})^{-1}$ of the electrons in the gas. The elastic collision cross section of electrons in He gas is $\sigma_{coll} \approx 7 \cdot 10^{-16}$ cm$^2$ [38] at electron energies $E_e \leq 4$ eV. For each gas sample, the helium number density $n_{He}$ is constant and given by the cell filling pressure $p_{He}$ and filling temperature $T_0$ ($T_0 \approx 300\,K$). Thus, the ratio of mean free path $\lambda$ to cyclotron radius $r_c$ weakly depends on the actual gas temperature *T* and essentially varies with the magnetic field strength *B* according to:

$$\frac{\lambda}{r_c} = \frac{\omega_c}{\nu_e} \approx 67 \cdot \frac{T}{T_0} \cdot \frac{B\,[\text{Tesla}]}{p_{\text{He}}\,[\text{mbar}]}. \tag{B4}$$

In our experimental conditions ($B = 4.7$ Tesla, $1 \leq p_{\text{He}}\,[\text{mbar}] \leq 15$) we find that $(\omega_c/\nu_e)^2$ always exceeds 440, the base value being obtained for the highest filling pressure. The primary effect of the magnetic field is to confine electrons within a cyclotron radius (of order µm at field strength of a few Tesla) and depress diffusion across the magnetic field whilst electrons can freely stream in the direction of *B* (*z*-axis). Moreover, electrons cannot pick up energy by the transverse component of the electric field apart from the negligible amount associated with the slow transverse magnetron motion at velocity $\upsilon = E \times B / B^2$. Hence transverse momentum can only be obtained indirectly from longitudinal momentum by collisions and remains necessarily much smaller on average. Therefore, in a magnetized plasma the momentum of the exciting electrons is fairly well aligned along *B* [10].

In the seminal paper by U. Fano and J.H. Macek [34] it is stated that the excitation of an atom or molecule by unidirectional collision in a gas leaves it generally in an anisotropic state. The collision process determines the components of the alignment tensor and orientation vector in a "collision frame" whose $\hat{z}$-axis usually coincides with the direction of an incident particle beam. In the simplest case, e.g., an unpolarized electron beam incident on gas atoms the experimental arrangement identifies only this $\hat{z}$-axis and has cylindrical symmetry about it. Under these circumstances, the alignment tensor has a single nonzero component (the situation met in the alignment experiment of [39], for example). The orientation vector, however, vanishes because it is an axial vector and no such quantity can be identified in a frame characterized by a single vector $\hat{z}$, unless the particles have nonzero helicity. For instance, optical pumping using circularly polarized light (incident photon beam) is the typical example of how one can directly achieve orientation in the atomic system.

In an aligned atomic system, states of different $|m_J|$ are populated unequally, while the populations in $m_J$ and $-m_J$ are the same. In contrast, an oriented system is characterized by differing populations in the $m_J$ and $-m_J$ states.

---

[10] The voltage drop across the rf discharge coil (cf. Fig 1b) causes an axial electric field (capacitively coupled discharge). Due to Faraday's law the axial magnetic field also generates an azimuthal electric field (inductively coupled discharge) which, however, is irrelevant for the gas discharge process because condition $\sigma_\perp \ll \sigma_\parallel$ is met.

Electron impact excitations of atoms and in particular the total cross sections for the $1^1S \to 2^3S$, $2^3P$ excitation from the $^3$He atomic ground state as well as the $2^3S \to 2^3P$, $3^3P$, $3^3D$, and $4^3D$ excitations from the metastable $2^3S$ state of helium have been calculated by [40]. At electron energies around 4 eV the $2^3S \to 2^3P$ collisional excitation peaks at $\sigma(2^3S \to 2^3P) \approx 10^{-14}$ cm$^2$, whereas the cross-sections of the other transitions are much smaller[11].

The collisional excitation rate for this transition is given by

$$\Gamma_{2^3S \to 2^3P} = n_e(4\,\text{eV}) \cdot \upsilon_e \cdot \sigma(2^3S \to 2^3P) \tag{B5}$$

The electron number densities $n_e$ at ~ 4eV reach values of $n_e > 10^{10}$ cm$^{-3}$ in medium and strong discharge plasmas [23,42,43] that results in $\Gamma_{2^3S \to 2^3P} > 10^4\,\text{s}^{-1}$. For comparison, the corresponding excitation rate from the atomic ground state to the $2^3P$ state[12] is $\Gamma_{1^1S \to 2^3P} \approx 10^{-3}\,\text{s}^{-1}$. Altogether, the creation rate (in cm$^{-3}$s$^{-1}$) of $2^3P$ atoms $\Gamma_{2^3S \to 2^3P} \cdot N^* + \Gamma_{1^1S \to 2^3P} \cdot N_{gs}$ is mostly set by excitations from the $2^3S$ state, although the proportion of $^3$He atoms in the $2^3S$ state ($N^*/N_{gs}$) lies in the 1 to 10 ppm range, at most, in our operating conditions [19]. According to the semiclassical approach introduced by Seaton [44] which is known as the impact parameter method, the $2^3S \to 2^3P$ collisional excitation cross section has the interpretation that, due to the field of the atomic electron, the colliding electron ($\hat{z}$-axis) emits a photon which is subsequently absorbed by the atom in the $2^3S \to 2^3P$ transition with $\omega_{SP} = 1.16$ eV/$\hbar$. The electric dipole matrix element of this transition corresponds to a $\pi$-transition ($\Delta m_L = 0$) which creates alignment in the excited state.

### B.2 Alignment-to-orientation conversion

Conversion of the excited-state alignment into orientation can occur during the time between excitation and decay. As shown by Fano and Macek [34], this cannot result from internal interactions alone, but can take place if these interactions (spin-orbit coupling or hyperfine-coupling, or both) are combined with the action of an external magnetic field (Zeeman-energy) which introduces the necessary axial vector into the Hamiltonian. Ignoring hyperfine terms (Hfs) for the present, the alignment-to-orientation conversion is most efficient in all cases when

$$\mu_B B(L_z + 2S_z) \approx a L \cdot S, \tag{B6}$$

---

[11] Cross sections for the spin-forbidden transitions He*($2^3S$)→He*($2^1P$), He*($2^1S$) can amount to $8 \cdot 10^{-16}$ cm$^2$ [41].

[12] $\Gamma_{1^1S \to 2^3P}$ can be derived from Eq. (B5), using $\sigma(1^1S \to 2^3P) \approx 5 \times 10^{-18}$ cm$^2$ [41], $n_e(25\,eV)/n_e(4\,eV) \approx 10^{-4}$ [23], and $\upsilon_e(25\,eV)/\upsilon_e(4\,eV) = 2.5$

i.e., when the Zeeman and spin-orbit energies are comparable [35] with *a* being the spin-orbit coupling strength.

Quantum mechanically, this AOC process may be viewed as resulting from the mixture of the wave functions of coupled fine-structure (Fs) levels and decoupled Paschen-Back levels in the intermediate field region where states with equal $m_J$ repel each other with opposite curvature, so called anti-crossings or avoided level crossings. As an important consequence there is a redistribution of the level's eigenfunctions and therefore of the populations near the avoided crossing point.[13]

Figure B1 shows the energy diagram of the $2\ ^3P_J$ sublevels of $^4$He – which we discuss first – as a function of the magnetic field strength. At zero field the spacings are due to the spin orbit interaction plus a marked downshift of the $2^3P_1$ state by residual interaction with the $2^1P_1$ state. The wave functions are well described by the coupled $|LSJ, m_J\rangle$ representation. At field strength B > 3 Tesla they are essentially decoupled in the $|LS, m_L, m_S, m_J\rangle$ representation. AOC occurs in the transition region between these two limiting cases where terms with the same $m_J$ repel each other through the spin orbit coupling and thus avoid the crossing.

F.W. Nehring [46] performed detailed numerical computations of the time averaged AOC coefficient $\langle F_0^{12} \rangle_t$ of the $2\ ^3P_J$ multiplet of $^4$He and identified two field regions where $\langle F_0^{12} \rangle_t$ peaks, with opposite signs: at low field, around $B \sim 0.2$ Tesla, where the $J = 2$ and $J = 1$ levels with the same $m_J$ quantum number repel each other and at high field, around $B \sim 1.1$ Tesla, where $\mu_B B$ is comparable to the 30 GHz splitting between the $^3P_1$ and $^3P_0$ states. The calculated magnetic-field dependence of the time-averaged orientation of the $2\ ^3P$ state of helium (following an instantaneous alignment) is shown by the black solid line in Fig. B2. This line reproduces the shape of the experimental polarization curve of McCall and Carver [17] and the zero crossing.

On a more elementary basis without advanced formalism, the transient AOC effect can be described using the analytical formula of eq. (B7) from Kemp et al. [35] derived for a simpler prototype case, i.e., an alkali-like atom or ion (without Hfs) with *s* ground-state and excitation into an aligned *p*-doublet state. If the radiative lifetime of the *p*-state is long compared to the spin-orbit and Zeeman precession times, the magnetic field dependence of the time-averaged fractional orientation, here denoted by $\langle q_a \rangle_t$, is given by

$$\langle q_a \rangle_t = \frac{-(a/\mu_B) \cdot B}{9/4 \cdot (a/\mu_B)^2 + B^2}, \tag{B7}$$

which reaches a deep minimum of -1/3 when $\mu_B B$ equals the doublet-splitting $(3/2)a$ and shows an asymptotic slope $\sim 1/B$. Applying tentatively eq. (B7) to the $2^3P_J$-triplet of $^4$He, we may set $a_1/\mu_B = 0.11$ Tesla and $a_2/\mu_B = 0.73$ Tesla to describe the low field maximum and the high field minimum, respectively, reported in Ref. [17]. The $\langle F_0^{12} \rangle_t$ curve apart from overall normalization and amplitudes' weighting factors is reproduced reasonably well (cf. Fig.B2, dashed black line) by

$$\langle F_0^{12} \rangle_t \approx \left|\langle q_{a_1} \rangle\right| - \left|\langle q_{a_2} \rangle\right|. \tag{B8}$$

---

[13] The effects of such situations are widely spread in physics. Related to NMR, anti-crossings provide the mechanism of HP transfer from an NMR silent state (singlet) to an observable state (triplet), e.g. [45].

Experimentally, we observe sizable nuclear polarization up to $P \sim -9\%$ at $B = 4.7$ Tesla, a field strength that lies well beyond the high field minimum and for which the predicted AOC coefficient does not exceed 2% (cf. Fig.B2). Even smaller polarization values may actually be expected for $^3$He ground state atoms, since the rf gas discharge plausibly induces partial atomic alignment only.

The obvious discrepancy between theoretical expectations at high field (from [46]) and our polarization measurements is a matter of investigation – in particular, also, with regard to the results of [35] on alkali-like atoms: the computed coefficient describing the evolution of aligned p-states into oriented ones can reach $\langle q \rangle = -33\%$ (cf. eq. (B7)).

It should be noted that Nehring [46] and Kemp et al. [35] primarily treat an astrophysical observation that was still unexplained then, namely the alignment and circular polarization of He-lines (predominantly $^4$He) occurring in sunspots which actually feature a magnetized plasma.

To include hyperfine-coupling ($AI \cdot J$) in going to $^3$He, additional contributions to $\langle F_0^{12} \rangle_t$ may occur when $g_J \mu_B B \cdot J \approx AI \cdot J$ where $g_J$ is the Landé g-factor for the $J$ multiplet. The induced orientation process can operate through the hyperfine and electron-Zeeman couplings in a manner very similar to the purely electronic process. In $^3$He the fine- and hyperfine interaction strengths are comparable, at least for the $J = 1$ and $J = 2$ states of the $2\,^3P_J$ multiplet. The hyperfine splitting yields 5 states $|LSJF, m_F\rangle$ with 18 Zeeman components altogether (cf. Fig. B1, left hand side). The present qualitative treatment cannot predict on a quantitative basis the presumable field-dependence of the AOC-effect induced nuclear ground state polarization of $^3$He including the metastability exchange mechanism. Regarding the Carver et al. data (cf. Fig. B2, blue data points), the inclusion of hyperfine coupling may account for the observed fine structure in the polarization curve measured around $\sim 0.2$ Tesla. An appropriate calculation of the AOC-induced nuclear polarization of $^3$He requires a full, field dependent diagonalization of the Fs- and Hfs-Hamiltonian as well as a proper treatment of the dynamics of collisional excitation and metastability exchange. This work is in progress and will be subject of a forthcoming publication.

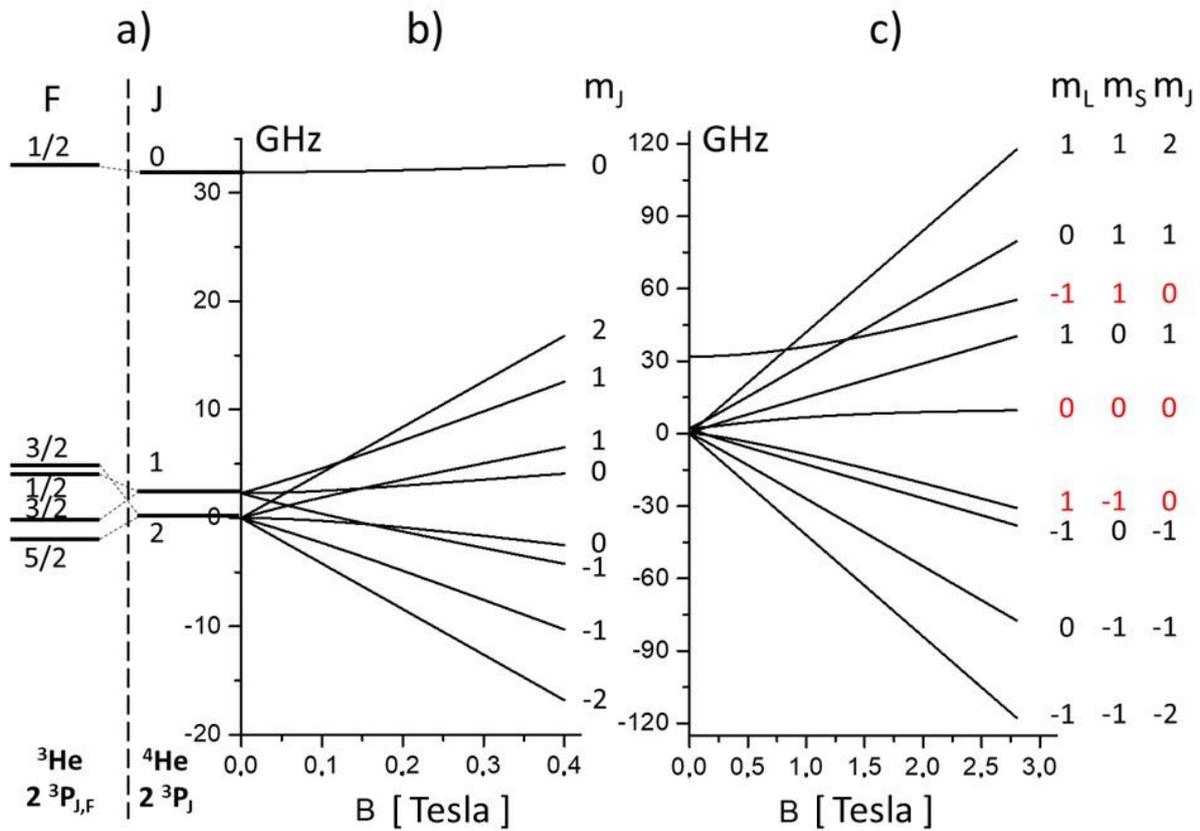

Fig. B1: a) Level structure in zero field of the $2^3$P-multiplets for $^3$He and $^4$He (isotope shift discarded). b) Zeeman-splitting of $^4$He states up to 0.4 Tesla. c) Zeeman splitting of $^4$He states up to 3 Tesla. On the right asymptotic Paschen-Back quantum numbers. Two distinguished field regions where an increased AOC occurs are around $B \sim 0.2$ Tesla where the $J = 2$ and $J = 1$ levels with the same $m_J$ quantum repel each other (anti-crossing) and at higher fields around $B \sim 1.1$ Tesla where the three ($m_J = 0$)-levels are involved in the high field avoided crossing.

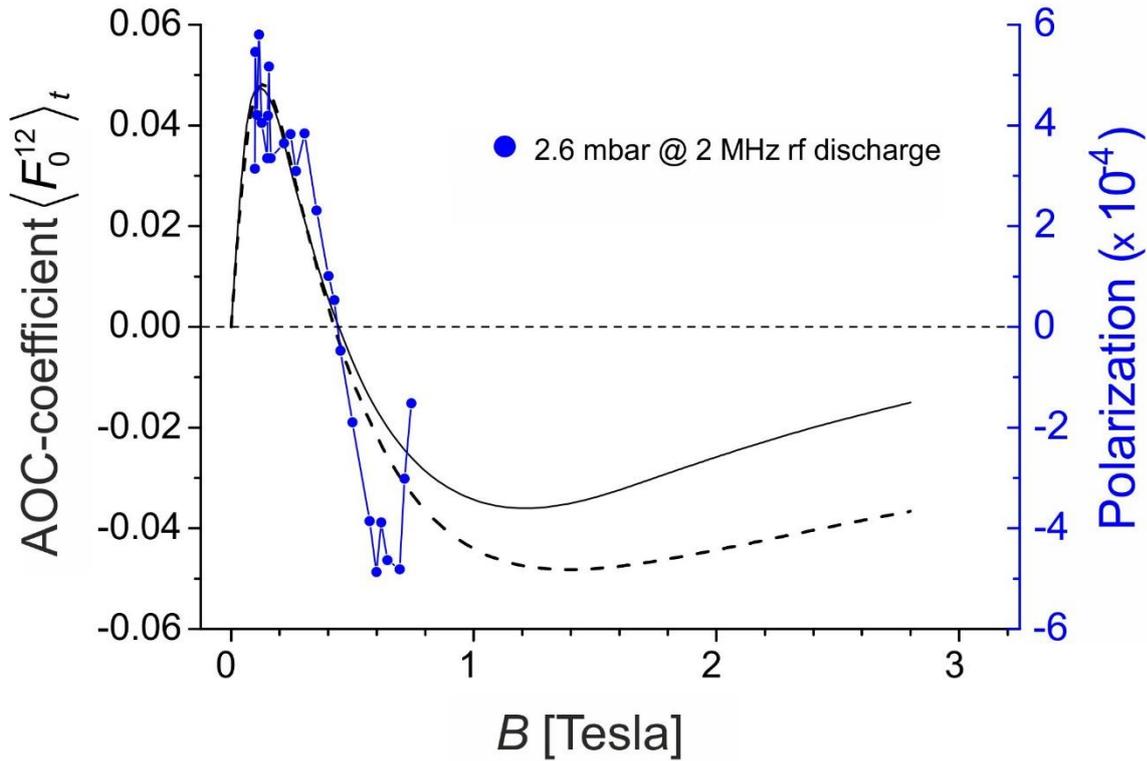

Fig. B2: Discharge nuclear polarization measured by McCall and Carver (blue data points, right vertical axis) as a function of magnetic field for a sample pressure of $p_{He} \approx 2.5$ mbar [17]. For $B > 0.7$ Tesla the drop in polarization magnitude was attributed to field inhomogeneities. Time-averaged AOC coefficient $\langle F_0^{12} \rangle_t$ (black solid line) for 2 $^3$P levels of helium ($^4$He) vs. the magnetic field strength as calculated by Nehring [46] using the formalisms of Fano and Macek [34]. Apart from a still unknown experiment-related conversion factor these theoretical results describe the field dependence of the measured polarization values with the sign change at $B \sim 0.45$ Tesla. The dashed line is a simplified analytical description of the field dependence of $\langle F_0^{12} \rangle_t$ according to eq. (B8) normalized to the first peak maximum of Nehring's result.